\documentclass{ws-procs9x6}

\newcommand{\kt}{\ensuremath{{\bf k}_{\perp}}}
\newcommand{\lt}{\ensuremath{{\bf l}_{\perp}}}
\newcommand{\mt}{\ensuremath{{\bf m}_{\perp}}}
\newcommand{\qt}{\ensuremath{{\bf q}_{\perp}}}
\def\bra#1{\langle{#1}|}
\def\ket#1{|{#1}\rangle}
\begin{document}

\title{Separation of Soft and Hard Physics in DVCS}

\author{A. G{\AA}rdestig, A.P. Szczepaniak and J.T. Londergan}

\address{Indiana University Nuclear Theory Center, \\
2401 Milo B. Sampson Lane, \\ 
Bloomington, IN 47408, USA\\ 
E-mail: agardest@indiana.edu}

\maketitle

\abstracts{ A model for deeply virtual Compton scattering, based on analytical
light-cone hadron wave functions is presented and studied at energies 
currently accessible at Jefferson Laboratory and DESY. 
It is shown that poles and perpendicular vector components play an 
important role at $Q^2<10$~GeV$^2$. A $Q^2$ suppressed diagram has to be
included at these low energies, but becomes negligible above 10~GeV$^2$.
Future prospects and developments of this model are discussed.}

\section{Introduction}
For many years, deeply inelastic scattering (DIS) has been a major source 
of our knowledge of parton distributions in nucleons and nuclei. 
This is because it can be shown that, at sufficiently high energies
and momentum transfers, the amplitude for this process factorizes 
into a `hard' part which can be calculated from QCD, and a `soft' 
part which can be extracted from experimental data. The soft part can 
be proved to be related to the probability of finding a quark with 
a particular flavor carrying a given fraction of the nucleon's 
momentum. 

In recent years, much interest has been focused on studying deeply 
virtual Compton scattering and the electroproduction of mesons, following the 
proof by Collins, Frankfurt and Strikman\cite{factor}. This proof demonstrated 
that, under quite general conditions, the leading amplitude for hard exclusive 
photo-production of mesons could also be factorized 
into a calculable `hard' part, and a `soft' part. All other amplitudes were 
smaller than the leading amplitude by powers of $1/Q$.  The soft part 
corresponds to the process by which a parton with a certain momentum fraction 
is removed from a nucleon, and replaced by a parton with a different momentum
fraction.  These parton distributions have been given a variety of names, but 
we will refer to them as `generalized parton distributions' (GPD's). 
In the case of quark helicity conservation, there are four independent GPD's;
$H(x,\zeta,t)$, $E(x,\zeta,t)$, $\widetilde{H}(x,\zeta,t)$, and 
$\widetilde{E}(x,\zeta,t)$, where $x$ and $\zeta$ are the light-cone momentum 
fractions of the struck quark and real photon, while $t=\Delta^2$ is the 
momentum transfer squared. Three physically different regions could be 
distinguished for $x$ and $\zeta$. 
The domain $0<\zeta<x<1$ ($\zeta-1<x<0$) corresponds to the removal 
and replacement of a quark (antiquark) with momentum fractions $x$ ($\zeta-x$)
and $x-\zeta$ ($-x$), respectively. In the remaining region $0<x<\zeta$, the 
photon scatters on a virtual quark-antiquark pair, extracted from the proton.
With this notation, $\zeta\to x_B$ (Bjorken $x$) in the limit $Q^2\to\infty$ 
($\Delta$ fixed).
In the limit of forward scattering (DIS), the $H$'s 
reduce to the quark density and quark helicity distributions:
\begin{eqnarray}
	H(x,0,0) & = & q(x) \\
	\widetilde{H}(x,0,0) & = & \Delta q(x).
\end{eqnarray}
The $E$'s do not appear in DIS, they are unique to the off-forward exclusive 
processes and provide information not accessible through other means. The
GPD's are related to the nucleon form factors by the integrals
\begin{eqnarray}
 \int_{\zeta-1}^1\frac{dx}{1-\frac{\zeta}{2}} H^q(x,\zeta,t) & = & F^q_1(t), \\
 \int_{\zeta-1}^1\frac{dx}{1-\frac{\zeta}{2}} E^q(x,\zeta,t) & = & F^q_2(t), 
\end{eqnarray}
and similarly for $\widetilde{H}$ and $\widetilde{E}$, all independent of 
$\zeta$. The factor $1-\frac{\zeta}{2}$ is included to comply
with the normalization of form factors used by Ji\cite{ji}.
The GPD's might also shed some light on the nucleon spin decomposition,
since Ji\cite{ji} has related them to the quark spin: 
\begin{equation}
	J_q = \frac{1}{2}\int_{\zeta-1}^1\frac{dx}{1-\frac{\zeta}{2}}\,
		x[H^q(x,\zeta,t=0)+E^q(x,\zeta,t=0)].
\end{equation}
The quark angular momentum is decomposed into $J_q=S_q+L_q$, where 
$S_q$ is measured in polarized DIS experiments. If the above sum-rule is 
measured in DVCS, the quark spin and orbital angular momentum parts could be 
separated. 

An alternative approach to DVCS is offered by the light-cone quark wave 
functions as suggested by Brodsky, Diehl, and Hwang\cite{BDH} (BDH). They have 
used this idea to calculate DVCS on an electron in QED for large $Q^2$.

There are ambitious efforts under way to measure DVCS (and the hard exclusive 
meson photo-production) at a number of facilities, DESY and JLab in 
particular\cite{exp}. At JLab energies, the competing Bremsstrahlung or 
Bethe-Heitler (BH) process is larger than DVCS. However, by doing interference 
measurements ($e^+/e^-$ beam charge asymmetry and various spin asymmetries) the
BH amplitude cancels out and only a BH$\times$DVCS interference remains. At 
DESY the energy is large enough for DVCS to become larger than BH, though the
measurements have low statistics, not allowing for differential cross sections.
Data for the asymmetries are, however, at both facilities collected for 
relative low beam energies and momentum transfers. 
It is thus interesting to investigate to what extent the `leading' amplitude 
is actually dominating in this kinematic region. We will here present the 
first few steps toward such an understanding, by developing a simple model 
using effective analytic quark-diquark wave functions, which allows us to 
study the amplitudes excited at these low energies. This model is similar to 
the one of BDH, but we calculate DVCS on a proton and focus 
especially on the features at low $Q^2$, keeping higher twist terms.

\section{Formalism}
\begin{figure}[t]
\centerline{\epsfxsize=4.5in\epsfbox{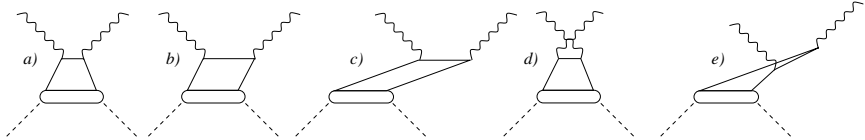}}   
\caption{The five different time-ordered `handbag' diagrams.}
\label{Fig:diag}
\end{figure}

We start from the Fourier transform of the $\gamma^{\ast}p\to\gamma p'$ 
amplitude (in light-cone coordinates)
\begin{equation}
       T^{++} = \int d^4y e^{iq'\cdot y}\bra{p'}TJ^{+}(y)J^{+}(0)\ket{p},
\end{equation}
where $J^{+}(y)=\bar\psi(y)\gamma^+\gamma_5\frac{\tau}{\sqrt{2}}\psi(y)$ is 
the electromagnetic current and $p(p')$ and $q(q')$ are the four-momenta of the
initial (final) hadron and photon. This expression could be expanded to give 
the five different one-loop diagrams (assuming light-cone gauge $q^+=0$) shown 
in Fig.~\ref{Fig:diag}. 
These diagrams are essentially the same as those used by BDH for DVCS on an 
electron.
The diagrams are represented by the integrals (because of space limitations, 
only $a$ and $b$ are given)
\begin{eqnarray}
T^{++}_a & = & 2(p^+)^2\int_{\zeta<x<1} \frac{dxd^2\kt}{16\pi^3}
	\phi^{\dagger}(z,\lt)
	\nonumber \\ & \times & 
     \frac{i}{\left[M^2+\left(\frac{1}{x_B}-1\right)Q^2-\frac{m^2+{\mt}_a^2}{x}
		-\frac{m_R^2+{\mt}_a^2}{1-x}+i\epsilon\right]}
	\phi(x,\kt), \\
\label{eq:TDef} 
T^{++}_b & = & 2(p^+)^2\int_{0<x<\zeta} \frac{dxd^2\kt}{16\pi^3}
	\frac{\left(m_R^2-\frac{m^2+\lt'^2}{z'}-
		\frac{M^2+\lt'^2}{1-z'}\right)
	\varphi(z',\lt')}
	{-\frac{\zeta(m^2+{\mt}_b^2)}{x(\zeta-x)}}
	\nonumber \\ & \times &
     \frac{i}{\left[M^2+\left(\frac{1}{x_B}-1\right)Q^2-\frac{m^2+{\mt}_a^2}{x}
		-\frac{m_R^2+{\mt}_a^2}{1-x}+i\epsilon\right]}
	\phi(x,\kt), 
\end{eqnarray}
where $z=(x-\zeta)/(1-\zeta)$, $z'=(\zeta-x)/(1-x)$, and the relative momenta 
are defined as $\lt=\kt+(1-z)\Delta_{\perp}$, $\lt'=-(1-z')\kt-\Delta_{\perp}$,
${\mt}_a=\kt+(1-x)\qt$, and 
${\mt}_b=\kt+\frac{x}{\zeta}\Delta_{\perp}+\frac{\zeta-x}{\zeta}\qt$. Here 
$x=k^+/p^+>0$ and $\zeta=q'^+/p^+$ are the longitudinal momentum
fractions of the struck quark and the real photon, $m$, $m_R$, and $M$ the 
quark, remnant (diquark), and hadron masses, and $\Delta = q-q'$. 
The expressions for the other diagrams have similar structures. 
In the limit $Q^2\to\infty$, the denominator of Eq.~(\ref{eq:TDef}) is 
proportional to $x-\zeta+i\epsilon$, i.e., the leading twist expression of 
Ji\cite{ji}. 
Note that the integrals are over different parts of phase space and include 
scattering on a quark or quark-antiquark pair only, not scattering on an 
antiquark. The wave function is represented by the analytical form
\begin{equation}
	\phi(x,\kt) = N\exp\left[-\frac{1}{\beta^2}
     \left(\frac{m^2}{x}+\frac{m_R^2}{1-x}+\frac{\kt^2}{x(1-x)}\right)\right],
\label{eq:wfdef}
\end{equation}
where $\beta=0.69$~GeV is chosen such that 
$F(q^2)=\int\frac{dxd^2\kt}{16\pi^3}\phi(x,(1-x)\qt+\kt)\phi(x,\kt)$
agrees with the dipole form factor for $Q^2<1$~GeV$^2$. 
The skew diagrams require the knowledge of the wave function 
$\phi(z',\lt')$ for the diquark splitting into a hadron and quark (lower 
right-hand corner of diagrams $b$, $c$, and $e$). 
The form of this wave function will eventually be restricted by exclusive data,
but is here arbitrarily chosen to be of the form of Eq.~(\ref{eq:wfdef}), with 
$m_R$ and $x$ replaced by $M$ and $z'$.

\section{Results}
\begin{figure}[t]
\vspace*{-2mm}
\centerline{\epsfxsize=3.0in\epsfbox{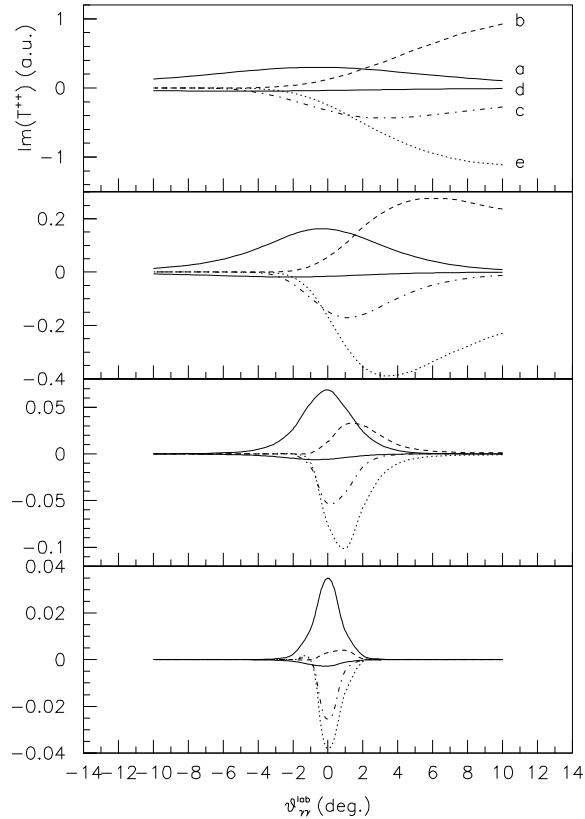}}   
\caption{The ${\rm Im}T^{++}$ matrix elements as functions
of the laboratory angle between the virtual and the real photon for in-plane
scattering. The labels correspond to the ones of Fig.~\protect\ref{Fig:diag}
(the lower solid line always correspond to diagram $d$).
The calculations are for $x_B=0.35$ and $Q^2=2,4,10,20$~GeV$^2$ 
from top to bottom.}
\label{Fig:Tpp}
\end{figure}

The $T^{++}$ matrix elements have been calculated for present JLab and DESY
kinematics and are shown in Fig.~\ref{Fig:Tpp}.
The laboratory angle $\theta_{\gamma\gamma'}$ between the virtual and real 
photons is defined for in-plane kinematics such that it is positive for 
$\phi=0$ and negative for $\phi=180^{\circ}$, where $\phi$ is the azimuth
angle between the final electron and proton, with ${\bf\hat{q}}$ as the polar 
axis.
These calculations incorporates the principal value parts of the integrals, 
i.e., the imaginary part of the $T^{++}$. 

The forward peaking of the skew diagrams ($b$, $c$, and $e$) is a consequence 
of the shift of the momentum $\kt$ in the extra denominator. The maximum of the
integral is dislocated to momenta where this denominator becomes very small. 

Because of its two hard propagator, diagram $b$ is expected to be suppressed in
the high $Q^2$ limit\cite{BDH}, and our calculation indicates that this 
suppression becomes significant for $Q^2>10$~GeV$^2$, i.e., well above present 
JLab energies.

While diagrams $a$ and $b$ exhibit a physical pole (for massless final photon),
the other diagrams never get vanishing denominators for physical values of 
kinematic parameters.
Thus the real part of the DVCS amplitude, which comes from the 
$\delta$-function part of the propagators, has contributions from these first
two diagrams only. The real part calculations will not be presented here.

This difference in pole structure also explains why diagram $d$ is smaller than
diagram $a$, since $Q^2$ terms add up in the denominator of $d$, while they
could cancel for $a$. This feature is closely related to the behavior of 
the two leading-twist propagators of Ji\cite{ji}; $1/(x-\xi+i\epsilon)$ and 
$1/(x+\xi-i\epsilon)$, where $x$ and $\xi$ are momentum fractions related to 
$\frac{1}{2}(p^++p'^+)$ instead of $p^+$. In this notation there is a
cancellation for $x=\xi$ (scattering on quark) or $x=-\xi$ (antiquark).
These simplified propagators neglect four-vector components, e.g., 
$\Delta_{\perp}$, that do not give large scalars in the Bjorken limit.


\vspace*{-1mm}
\section{Conclusions and outlook}
We have introduced a model for DVCS, using simple quark wave functions and 
retaining all components of four-vectors throughout. The perpendicular momenta
are shown to be very important for the location of the maxima of the integrals 
at low $Q^2$.
Within the model we have been able to investigate the five single-loop 
diagrams and their relative importance for DVCS at JLab and HERA energies.
The results indicate that at present JLab energies ($Q^2<4$~GeV$^2$), all of 
the diagrams (except possibly the crossed diagram $d$) need to be considered. 
In particular it is necessary to include diagram $b$, despite its two hard 
propagators, since the expected suppression is significant only for larger 
$Q^2$. Other higher-order diagrams might need to be considered as well.
At $Q^2>10$~GeV$^2$, the process is completely dominated by the handbag 
diagrams with one hard propagator. 

This work will be extended to cover a wider kinematic range in $Q^2$ and $x_B$
and will include antiquark contributions to the amplitudes and the wave 
functions. The photo-production of mesons will also be calculated, with special
consideration of the meson poles then made possible in the skew diagrams. 

In order to compare with actual JLab and HERA experiments the interference with
the Bremsstrahlung (Bethe-Heitler) process has to be calculated. 
We would then obtain cross sections and asymmetries to test against 
experimental results and be able to check the  validity of commonly used 
approximations and to evaluate various wave functions that could be used.

This work was supported in part by NSF grant NSF-PHY0070368 and DOE grant
DE-FG02-87ER40365.


\vspace*{-2mm}
\begin{thebibliography}{0}
\bibitem{factor} J.C. Collins, L. Frankfurt and M. Strikman, {\it Phys.\ Rev.} 
	{\bf D56}, 2982 (1997).

\bibitem{ji} X. Ji, {\it Phys.\ Rev.\ Lett.} {\bf 78}, 610 (1997); 
	{\it Phys.\ Rev.} {\bf D55}, 7114 (1997).

\bibitem{BDH}
S.~J.~Brodsky, M.~Diehl and D.~S.~Hwang, Nucl.\ Phys.\ {\bf B596}, 99 (2001).

\bibitem{exp} see e.g., A. Airapetian {\it et al.}, 
	{\it Phys.\ Rev.\ Lett.} {\bf 87}, 182001 (2001); 
  S. Stepanyan {\it et al.}, {\it Phys.\ Rev.\ Lett.} {\bf 87}, 182002 (2001).

\end{thebibliography}
\end{document}